\documentclass[%
 reprint,
superscriptaddress,
 amsmath,amssymb,
 aps,
]{revtex4-2}

\usepackage{color}
\usepackage{amsthm}
\usepackage{graphicx}
\usepackage{dcolumn}
\usepackage{bm}
\usepackage{mathrsfs}
\usepackage{dsfont}
\usepackage{selinput}
\usepackage{mathtools}

\newcommand{\md}{\mbox{d}}

\definecolor{lana}{RGB}{148,0,211}
\definecolor{gsr}{rgb}{0.53, 0.15, 0.34}

\begin{document}

\preprint{APS/123-ART}

\title{Quantum probing of null-singularities}  

\author{Ivo Sachs}%
 \email{ivo.sachs@physik.uni-muenchen.de}
 \affiliation{Arnold Sommerfeld Center for Theoretical Physics, Theresienstra{\ss}e 37, 80333 M\"unchen, Germany\\}
 \author{Marc Schneider}%
 \email{marc.schneider@aei.mpg.de}
 \affiliation{Albert-Einstein Institute for Gravitational Physics, Am M\"uhlenberg 1, 14476 Potsdam, 
 Germany\\}
 \author{Maximilian Urban}%
 \email{maximilian.urban@physik.uni-muenchen.de}
\affiliation{Arnold Sommerfeld Center for Theoretical Physics, Theresienstra{\ss}e 37, 80333 M\"unchen, Germany\\}

\date{\today}

\begin{abstract}
We adapt the dual-null foliation to the functional Schr\"odinger representation of quantum field theory
and study the behavior of quantum probes in plane-wave space-times near the null-singularity.
A comparison between the Einstein-Rosen and the Brinkmann patch, where the latter extends beyond the first,
shows a seeming tension that can be resolved by comparing the configuration spaces. Our analysis concludes
that Einstein-Rosen space-times support exclusively configurations with non-empty gravitational memory
that are focused to a set of measure zero in the focal plane with respect to a Brinkmann observer. To conclude, we provide a rough framework to estimate the qualitative influence of back-reactions on these results.

 \end{abstract}

 \maketitle
\section{Introduction} 
Space-times featuring plane-fronted waves with parallel rays (pp-wave space-times)
represent an important class of exact solutions to the Einstein equations as they 
describe non-linear gravitational waves in general relativity. The main applications of these space-times 
can be found in string theory, as these backgrounds support exact string solutions in scattering theory
as an effective description of a non-perturbative scattering, or in terms of the gravitational 
memory effect \cite{zel74} which will play a role later in this article. One special subclass thereof 
are plane-wave space-times which portray the even simpler situation where the profile of the wavefront is
constant along the transversal direction. Penrose extensively studied these space-times
from a geometric perspective in a series of seminal articles \cite{pen65, pen76}. Those space-times' most 
distinguished property is the focusing of null-rays that have crossed the wave. Since the
weak-energy condition holds, the plane wave will act like a converging lense such that collimated light-rays
meet in a focal plane. The space-time then develops two focal planes with respect to future and past directed 
null-rays. This intersection of null-geodesics indicates that the manifold itself is null-geodesically 
incomplete. A further consequence is that this class does not admit a Cauchy surface and therefore fails to be 
globally hyperbolic. Albeit curvature invariants are vanishing globally, null and timelike 
geodesic incompleteness is the minimal criterion for a space-time to be called singular \cite{hawk73}.

Owing to their peculiar causal structure, plane-wave space-times can be described by so-called 
Einstein-Rosen patches, i.e. coordinate systems
ranging from null infinity to the focal plane, where the metric degenerates. 
Since focusing singularities are classified as weak, one can still find a consistent
extension of the Einstein-Rosen 
patches beyond the singularity, subject to certain integrability conditions \cite{el77}. This extension, 
the Brinkmann patch, fully covers the 
manifold from past to future null-infinity without degenerating. 

In recent analyses 
\cite{hof15,hof17,egl17,hof19,jur18,di19},
classical singularities have been probed by quantum fields within the functional Schr\"odinger formalism
with intriguing results, e.g. Schwarzschild black-holes admit a consistent quantum field theory. Plane fronted 
waves provide an excellent setting to further investigate the concepts developed therein, as they are distinctly 
different from previous examples. As the singularity is 
null and therefore of a different type from previous studies it may shed some light on how the nature of the 
singular surface, as well as the bordering space-time, impacts previously seen effects. Furthermore, these space-times have zero curvature 
everywhere away from the wave-front. In fact, the regions flanking the wave are Minkowski patches which allows us to use properties of flat 
space-time.
In particular, one does not expect to observe
dissipative effects, such as those occurring in 
Schwarzschild space-time. Hence, 
a focused massless quantum field theory is predicted to encounter the null-singularity, providing a 
consistency check for the previous interpretation of quantum completeness.

The non-existence of Cauchy hypersurfaces imposes an obstacle to the definition of a sensible evolution 
problem, as this strongly depends on a well-defined initial-value problem and Hamiltonian flow. A remedy was 
found by Hayward\cite{ha93, hay93, hayw93},
making use of the fact that well-defined initial value problems prohibit only time-like
separations between two points on the initial hypersurface. The resulting dual-null foliation allows for the
construction of Lagrangian and Hamiltonian dynamics. These then support well-defined initial value problems,
as well as Hamiltonian flows, such that the Cauchy problem can be generalized to some non-globally hyperbolic
space-times. The construction also admits a consistent path integral quantization,
making it particularly useful.

In this article, we utilize the dual-null foliation to derive the functional Schr\"odinger 
representation of quantum field theory in these cases. While our construction is completely general and 
applicable to all space-times, the particular geometry of plane-wave space-times makes the application 
especially simple and transparent.

In the following section, we
review some geometrical features of plane-wave space-times
with the emphasis on the different coordinate systems.
Section \ref{20} briefly discusses Hayward's dual-null foliation for Hamiltonian developments in general 
before it is explicitly applied to the functional Schr\"odinger analysis. After constructing the Hamilton
operator and the corresponding states, we will perform a null-reduction that reflects our initial conditions.
In section \ref{IW}, the special case of a delta-distribution shaped shock-wave is examined explicitly in
both coordinate patches. We show that there appears to be a tension between the two results which can be 
resolved by carefully studying the underlying configuration spaces. 
The last section discusses back-reactions on the system. Similarly to \cite{sachs19}, we chose a
heuristic approach that yields a corrected metric around the focal plane and estimates its influence on 
the Schr\"odinger wave-functional. Afterwards, we set our results in context to the previous results in 
dynamical space-times and discuss further directions. Note that we will work in the units $c=\hbar=G_N=1$ 
throughout the article.

\section{Plane-wave space-time} 
The general four-dimensional space-time $(\mathbb{W}^4,g)$ describing a gravitational-electromagnetic
pp-wave is given by a semi-Riemannian manifold $\mathbb{W}^4$ and the corresponding 
metric in Brinkmann or harmonic coordinates \cite{pen65}
\begin{equation}\label{pwst}
g=-2\mbox{d}u\otimes\mbox{d}v+H(u,x)\md u\otimes\md u+\delta_{ab}\md x^a\otimes\md x^b,
\end{equation}
where the function $H(u,x)$ describes the profile of an outgoing plane-wave that propagates through 
Minkowski space. Here, $u$ is the ingoing and $v$ the outgoing null-direction while
the $x^a$'s are spatial coordinates with $\delta_{ab}$ being the identity matrix. This space-time can be 
thought of two as Minkowski patches linked by a plane-wave of compact support
in the $u$-direction (sandwich wave).
The pp-wave metric enjoys a high degree 
of symmetry and is therefore characterized almost entirely by $H$. For the special case of plane waves,
the function $H(u,x)$ is quadratic in $x^a$, that is, $H(u,x)\to H_{ab}(u)x^ax^b$.
In particular, gravitational waves are described by profile functions that fulfill tr$(H)=0$, i.e. have a 
vanishing Ricci tensor, while electromagnetic waves, $H_{ab}(u)=H(u)\delta_{ab}$ are characterized by 
vanishing Weyl tensors. The Ricci
tensor in this space-time is then non-trivial at all points through which the wave propagates, i.e. 
$R_{uu}=-\mbox{tr}(H)(u)$, while other components are zero. Hence, \eqref{pwst} is the solution to the
Einstein equation with Einstein tensor $G_{\mu\nu}=0$ except for $G_{uu}=8\pi T_{uu}=-\mbox{tr}(H)(u)$.
Focusing properties of plane waves are implied since the weak energy condition tr$(H)\le0$ holds. It follows 
from the fact that the profile function $H(u,x)$ is independent of $v$ that there are no self-focusing effects on the wavefront itself.
%
\begin{figure}
\includegraphics[width=6cm]{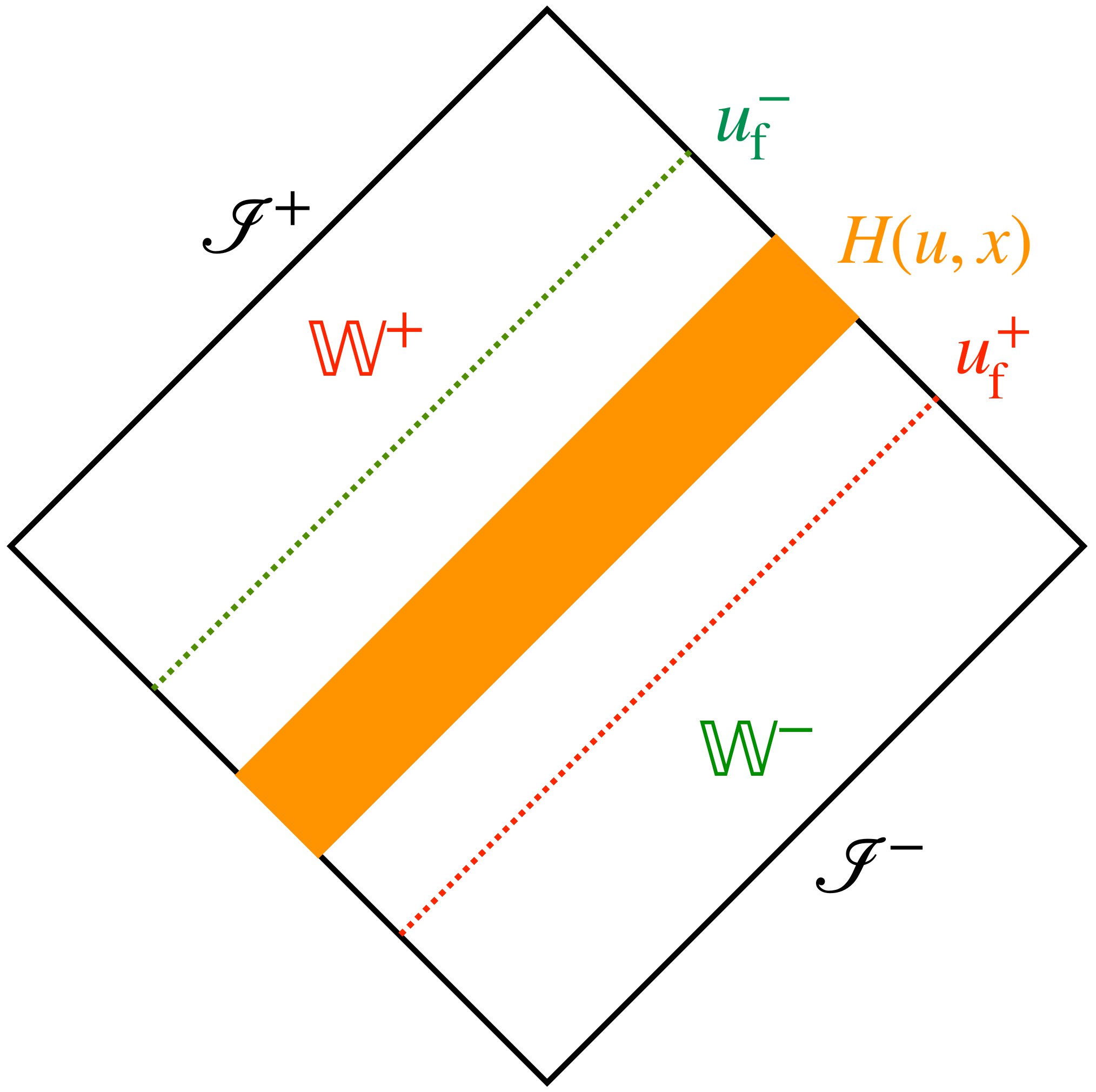}
\caption{
\label{fig:pen} 
The Carter-Penrose diagram of the pp-wave space-time shows the full Brinkmann chart $(\mathbb{W}^4,g)$,
with embedding $\iota:\mathscr{U}\times\mathscr{V}\times\mathcal{S}\to\mathcal{M}$, 
ranging from $\mathscr{I}^-$ to $\mathscr{I}^+$. The outgoing wave $H(u,x)$ is depicted by the
orange bar that divides the manifold into two parts, where all
white areas describe Minkowski space-time.  Additionally, we see
the two Einstein-Rosen patches $\mathbb{W}^-=(-\infty,u^-_{\rm f})\times\mathscr{V}\times
\mathcal{S}$ and $\mathbb{W}^-=(\infty,u^+_{\rm f})\times\mathscr{V}\times\mathcal{S}$ that 
end at the null-singularities at the focusing points $u^\pm_{\rm f}$ displayed by the dashed lines.}
\end{figure}

The first extensive study of these space-times was undertaken by Penrose \cite{pen65,pen76} and found that, 
although they are strongly causal, they do not admit Cauchy hypersurfaces due to a focusing singularity. 
More specifically, every light-cone will degenerate in at least one direction at some distance after crossing
the wave-front. Thus, all geodesics are destined to meet at said focal point $u_{\rm f}$ clearly excluding the
possibility of a Cauchy surface as defined in \cite{pen65}.

While the Brinkmann coordinates are useful as they cover the entire manifold, the symmetries of the space-
time are more transparent in the Einstein-Rosen or group coordinates
\begin{equation}\label{rosen}
	g=-2\mbox{d}U\otimes\mbox{d}V+\gamma_{ij}\mbox{d}y^i\otimes\mbox{d}y^j .
\end{equation}
The transition functions between \eqref{pwst} and \eqref{rosen} are given through the vierbein $E^a_i(u)$, which in this case is only a function of $u$
and satisfies the Einstein equation $\ddot{E}_{ai}=H_{ab}(u)E^b_i$, together with the 
relation $g_{ij}=E^a_iE^b_j\delta_{ab}$ and the symmetry condition 
$\dot{E}_{ai}E^i_b=\dot{E}_{bi}E^i_a$ stemming from the Wronski determinant. 
The asymptotic symmetries of Minkowski space-time must be reflected by the 
vierbeins such that $\lim_{u\to\pm\infty}E^\pm_{ia}={\delta}_{ia}$ \cite{neko17}.
Explicitly, the coordinate transformations that lead to \eqref{rosen} are given by: $u\to U$,
$v\to V+\tfrac{1}{2}\dot{\gamma}_{ij}(U)y^iy^j$, and $x^a\to E^a_iy^i$. The metric
of the two-dimensional spatial sub-manifold can be constructed from the vierbeins via
$\gamma_{ij}(U)=\tfrac{1}{2}(E^a_iE_{aj}+E^a_jE_{ai})$.
The over-dot unambiguously denotes a derivative with respect to the ingoing direction
$u$- or $U$-coordinate, as the two are identical.
In \eqref{rosen}, we can immediately read off the Killing vectors $\tfrac{\partial}{\partial V}$ and $\tfrac{\partial}{\partial y^i}$.
The remaining ones are given by the combination
$\mathcal{D}^i=y^i\tfrac{\partial}{\partial V}+F^{ij}(U)\tfrac{\partial}{\partial y^j}$ with 
$F^{ij}(U)=\smallint^U\md\upsilon\gamma^{ij}(\upsilon)$.
The Killing vectors $D$ in \eqref{pwst} can be constructed directly from the vierbeins $E$ and are 
$D^i=E^{i a}\tfrac{\partial}{\partial x^a}-{\dot{E}^i}_a x^a\tfrac{\partial}{\partial v}$.
Geometrically, these can be thought of as connecting vectors between neighboring geodesics \cite{garr91}. 

Although most calculations are easier in Einstein-Rosen coordinates, the drawback is
that they only cover a part of $\mathbb{W}^4$, either from past null-infinity $\mathscr{I}^-$ to the
focusing plane
in the future of the wave or from future null-infinity $\mathscr{I}^+$ to the focusing
singularity in the past of the wave. We will call the chart with the asymptotic boundary in the future  
$\mathbb{W}^+$ and 
the chart with the asymptotic boundary in the past $\mathbb{W}^-$. 
The geometry is illustrated in the Penrose diagram in FIG. \ref{fig:pen}.
Here, the Brinkmann patch describes the entire diamond while the Einstein-Rosen charts 
$(\mathbb{W}^\pm,g^\pm)$ range from $\mathscr{I}^\pm$ to the null-singularity at $u_{\rm f}^\pm$ and 
together
cover the whole space-time with overlap between the null-singularities.
It should be mentioned that
the distance in $u$, or $U$,
between the focal point and the wave-front is inversely proportional to the amplitude
of the wave. 
The Killing vectors alone already encode many of the interesting features of plane-wave space-times such as 
for example the gravitational memory effect. 
Null-geodesics starting at $\mathscr{I}^-$ experience a Minkowski evolution, that is to say, 
$E^-_{ia}=\delta_{ia}$ before they have crossed the wavefront. It is therefore clear that we recover the 
spatial translation invariance of Minkowski 
space-time by finding $D^a\to\tfrac{\partial}{\partial x^a}$. 
However, once a geodesic has crossed the wave, the 
vierbein develops a $u$-dependence that triggers the deviation of ingoing geodesics towards the 
$v$-direction resulting in a focusing. This $u$-dependence reflects the gravitational memory as it persists
throughout the entire future development.
In the patches $\mathbb{W}^-$ and $\mathbb{W}^+$ the spatial
metric accounts for the gravitational memory as $\gamma^\pm_{ij}\to\delta_{ij}$
at the asymptotic boundary $\mathscr{I}^\pm$, however it will develop a $U$-dependence that leads to a 
degeneracy at some $U^\pm_{\rm f}$ beyond the pp-wave. In other words, the spatial hypersurfaces in 
Einstein-Rosen coordinates
will degenerate based on the nature of the wave itself: while for electromagnetic waves, the leading order in 
the mode expansion is a dipole that supports an isotropic contraction, gravitational waves
are governed by quadrupole modes that contract in one direction and simultaneously 
expand in the perpendicular direction resulting in an astigmatic focusing 
(caustic).

From the transformation between the spatial coordinates $x_a$ and $y_i$ we see that the spatial volume of the
Einstein-Rosen patch collapses, while remaining constant for Brinkmann coordinates. The latter
is hence an extension of space-time through the singularity with a non-degenerating spatial part as can
be seen in \eqref{pwst}. This is possible because the null-singularity 
occurring here is weak according to
the definition of \cite{el77} such that a geometrical extension is possible. The 
gravitational memory is seen in the $u$-dependence of the $E_{ia}$ which strongly influences the
Killing vectors.

\section{Dual-null Foliation}\label{20}

The global hyperbolicity of space-time is considered essential for most analyses in Hamiltonian systems, 
however, generically this 
requirement is not necessary for a well-defined initial value problem. Following \cite{hawk73}, 
the necessary and 
sufficient condition is the existence of an achronal set (a surface without two events that are 
separated by a time-like curve) and a global flow which respects this property of the set. Hayward
constructed in a series of seminal articles \cite{ha93, hay93, hayw93}
the dual-null formulation of Hamiltonian dynamics, showing that 
the initial value problem of the Cauchy evolution from the initial Cauchy surface is replaced by a 
combined boundary and initial value problem in the null-evolution. More precisely, the initial data is prescribed
by two intersecting null-surfaces in terms of a boundary value problem on the null-infinities
and a set of initial values
on the intersection. This introduces a new set of canonical momenta, consistency is enforced through 
additional integrability conditions.

The first step is to create a suitable embedding:
let $\mathcal{M}$ be a $d$-dimensional, globally time-orientable mani\-fold, then we
construct a $(d-2)$-dimensional, compact, orientable and time-orientable,
space-like sub-manifold $\mathcal{S}$ \cite{hay94}. Additionally, we take two half-open intervals 
$\mathscr{V}=[0,v)$ and $\mathscr{U}=[0,u)$ such that we can define a smooth embedding 
$\iota:\mathscr{U}\times\mathscr{V}\times\mathcal{S}\to\mathcal{M}$. Using $\iota$
we can build the phase space of the elementary observables:
consider a vector bundle $Q$ over the base space $B$; then the 
configuration variable will be $q\in CQ$, where $C$ denotes the space of smooth sections. We can now define 
the two directions
given by vectors in $\mathscr{U}$ and $\mathscr{V}$ such that they are aligned with the null-congruences.
Hence, the space-time is foliated by two sets of three-surfaces $\Sigma_-$ along the $u$-direction and 
$\Sigma_+$ along the $v$-direction. The manifold is thus covered by two stacks of hypersurfaces along 
the two corresponding null directions.

\subsection{Hamilton density}
Consider the tangent space $TCQ$ to $CQ$, we can define the velocity fields 
$(q,q^+,q^-)\in TCQ\oplus TCQ$. From here it is clear that the evolution space is given by $\mathscr{V}
\times\mathscr{U}$ and $q^+$ is the velocity field tangent to the outgoing $v$-direction and 
$q^-$ is tangent to the ingoing $u$-direction. The Hamilton density can be constructed by a 
Legendre transformation \cite{hayw93} of the Lagrange density $\mathcal{L}$
\begin{equation}\label{hami}
\mathcal{H}(q,p^+,p^-)=((0,p^+,p^-)-\mathcal{L})\Lambda^{-1}(q,p^+,p^-).
\end{equation}
Here the conjugate fields are $p^\pm\in T^*CQ$ where $\Lambda:
TCQ\oplus TCQ\to T^*CQ\oplus T^*CQ$, 
$\Lambda: (q,q^+,q^-)\mapsto (q,\tfrac{\delta\mathcal{L}}{\delta q^+},
\tfrac{\delta\mathcal{L}}{\delta q^-})$ denotes the invertible Laplace transformation. The identification of the conjugate momenta $p^\pm$ with the 
functional derivative of
$\mathcal{L}$ with respect to the velocity fields $q^\pm$ follows from the Hamilton equations $p^+=
\tfrac{\delta\mathcal{L}}{\delta q^+}$,
$p^-=\tfrac{\delta\mathcal{L}}{\delta q^-}$, and $
\tfrac{\partial q}{\partial u}=\tfrac{\delta\mathcal{H}}{\delta p^+}$,
$\tfrac{\partial q}{\partial v}=\tfrac{\delta\mathcal{H}}{\delta p^-}$,
$\tfrac{\partial p^+}{\partial u}+\tfrac{\partial p^-}{\partial v}=-\tfrac{\delta\mathcal{H}}{\delta q}$ with 
the additional 
integrability condition $\tfrac{\partial}{\partial u}(\tfrac{\delta\mathcal{H}}{\delta p^-})=
\tfrac{\partial}{\partial v}(\tfrac{\delta\mathcal{H}}{\delta p^+})$. In the double null case, it was shown by 
Hayward that these can always be satisfied \cite{hay94}.
\begin{figure}
\includegraphics[width=7.5cm]{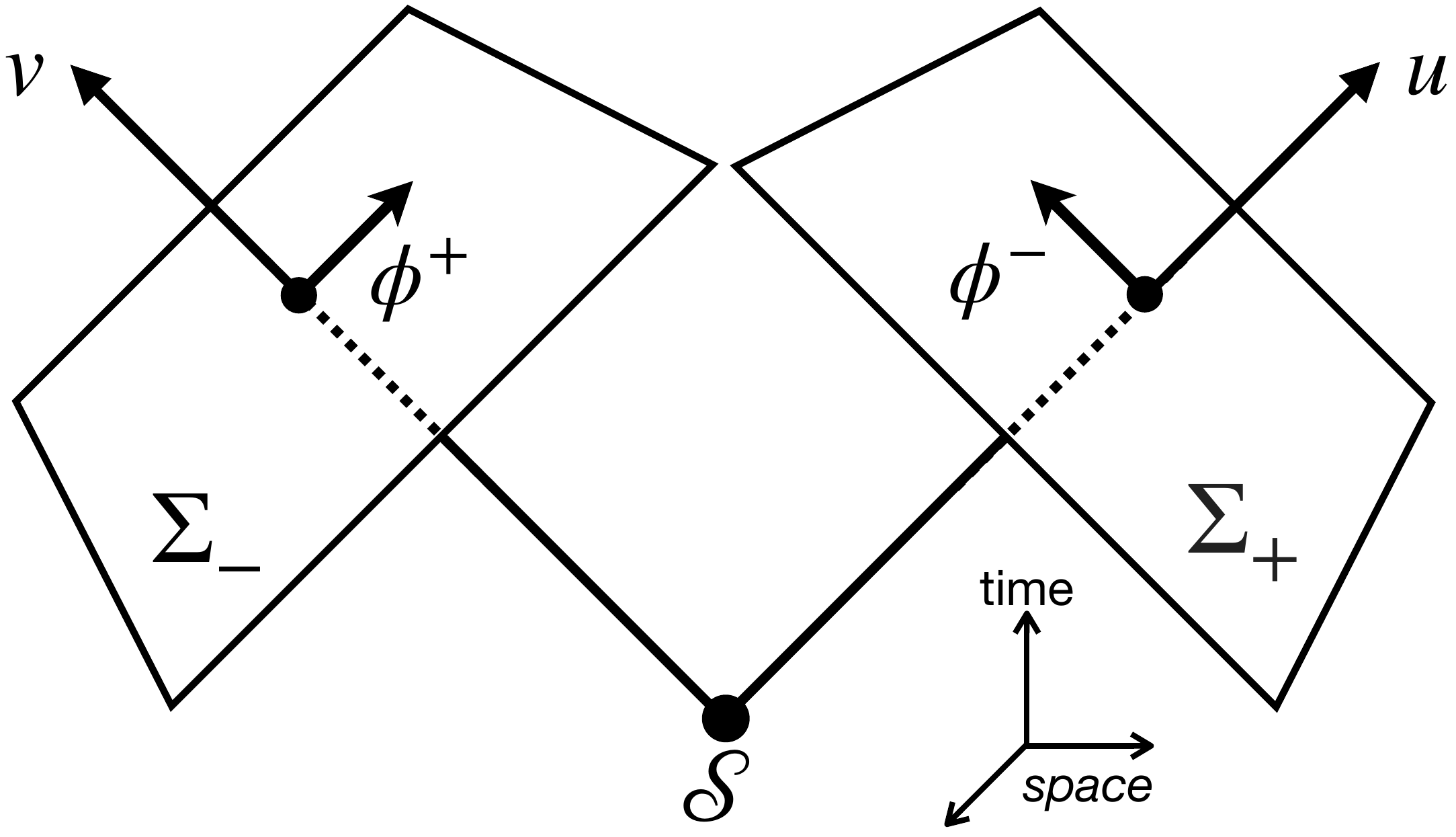}
\caption{
\label{fig:2al0} 
Sketch of the dual-null foliation (spatial dimensions partially suppressed):
 attached to the two dimensional space-like surface $\mathcal{S}$ on which 
the spatial fields live, are the 
ingoing and outgoing null-vectors along  $v\in\mathscr{V}$ and $u\in\mathscr{U}$ respectively.
Perpendicular to the null-directions are the three-surfaces $\Sigma_-$
and $\Sigma_+$ and the velocity fields $\phi^\pm\upharpoonright\Sigma_\mp$. Points in the 
two-dimensional spatial sub-manifold are represented by intersections of the three-submanifolds $\Sigma_+$
and $\Sigma_-$ at certain values of the light-cone coordinates.}
\end{figure}

The above construction can be applied directly to the quantization of fields in curved spaces where it is 
especially useful to study transmissions through null-surfaces.
Plane wave space-times particularly lend themselves to this setup, as there is a natural choice of null surfaces
dictated by the wave front. Hence, we consider a free massless scalar field theory 
$S=-\tfrac{1}{2}\textstyle{\int}{\rm d}^4x \sqrt{-g}\:(g(\nabla\phi,\nabla\phi)+\zeta\mathcal{R}\phi^2)$
with covariant derivative $\nabla$,
Ricci scalar $\mathcal{R}$, and coupling constant $\zeta$. Rewriting this action in Brinkmann
coordinates, using the embedding $\iota$ given by the dual-null foliation, yields
\begin{align}
S=-\frac{1}{2}\int_{\mathbb{W}^4}\md^4x\sqrt{-g}&\left(-2\phi^+\phi^-
+H(u,x)\phi^+\phi^+\right.\nonumber\\
&\left.+\delta(\nabla\phi,\nabla\phi)+\zeta\mathcal{R}\phi^2\right)\label{accion}
\end{align}
where $\delta$ is the induced Euclidean flat metric on $\mathcal{S}$ and the $\nabla$'s are 
understood to be the spatial derivatives. Considering $\iota$, 
the boundary and initial data for the velocity fields 
with respect to the $u$- and $v$-directions are given by $\phi^+\upharpoonright
\Sigma_-=\{0\}\times\mathscr{V}
\times\mathcal{S}$,  and $\phi^-\upharpoonright
\Sigma_+=\mathscr{U}\times\{0\}\times\mathcal{S}$, while the spatial data are given by
$\phi\upharpoonright\{0\}\times\{0\}\times\mathcal{S}$. 
An illustration of the dual-null foliation can be found in FIG. \ref{fig:2al0} where the
velocity fields and the hypersurfaces are shown explicitly. From here, 
it is obvious why the evolution in this foliation is an initial value problem on $\mathcal{S}$ combined with
a boundary condition at $\mathscr{I}^+$ or $\mathscr{I}^-$. 

Applying \eqref{hami} to \eqref{accion}, we find the Hamilton density in Brinkmann coordinates to be
\begin{equation}
\mathcal{H}=\frac{1}{2}\left(2\pi^+\pi^-+
H(u,x)\pi^-\pi^-
+\delta\left(\nabla\phi,\nabla\phi\right)\right).\label{fham}
\end{equation}
This is the general form for a Hamilton density in Brinkmann coordinates, however,
for an outgoing wave, the only non-zero
component of the Ricci tensor is $R_{uu}=-tr(H)$ which is zero for gravitational waves as we have 
seen before. Hence, we have safely set $\mathcal{R}\equiv0$ in the above equation. Furthermore, 
to replace the mixing terms in the Legendre transformation we use the Hamiltonian equations 
\begin{eqnarray}
\pi^-&=&\frac{\delta\mathcal{L}}{\delta\phi^-}=\phi^+,\label{leg1}\\
\pi^+&=&\frac{\delta\mathcal{L}}{\delta\phi^+}=\phi^--H(u,x)\phi^+\label{leg2}.
\end{eqnarray}
We see that the outgoing momentum is not affected by the wave since it propagates parallelly
while the ingoing
momentum that crosses the wave experiences a distortion which will lead to the inevitable focusing. Hence, 
the first term on the right hand side of \eqref{hami} becomes $\pi^-\phi^-=\phi^+\phi^-$ 
and $\pi^+\phi^+=\phi^-\phi^+-H(u,x)\phi^+\phi^+$ such that we can complete the 
Legendre transformation using \eqref{leg1} and \eqref{leg2} to get \eqref{fham}.
Note that the integrability condition in this space-time can be satisfied trivially because
it simply reduces to Schwartz's theorem for the exchange of partial derivatives. Although the
Hamilton density in Einstein-Rosen coordinates misses the term proportional to $H(u,x)$ and therefore 
speciously looks simpler, the spatial metric will be given by $\gamma$ and has a non-trivial $U$-dependence.

\subsection{Functional Schr\"odinger states}

In this part, we demonstrate how to construct the functional Schr\"odinger representation in the dual-null
foliation. For a Cauchy problem with time-direction $t$ the differential operator describing the
Schr\"odinger equation $\hat{P}\psi=0$ with Hamilton operator $\hat{H}(t)$ is given by 
$\hat{P}=i\partial_t-\hat{H}(t)$. To apply the above formalism to the Schr\"odinger representation, let
us recall the definition of the light-cone 
coordinates $u=t-z$ and $v=t+z$, with $z$ being a spatial direction. Hence, $\partial_t\to\tfrac12(
\partial_u+\partial_v)$ in this coordinate chart and the Schr\"odinger equation becomes
$\hat{P}=i(\partial_u+\partial_v)-2\hat{H}(u,v)$ accordingly. We see that the evolution equation in this foliation
looks involved due to the combination of evolution directions. In the functional Schr\"odinger representation,
the differential operator will be constructed using \eqref{fham}. The difference from a 
quantum-mechanical state $\psi$ is that in this representation the 
states are wave-functionals $\Psi$ that read in
the instantaneous fields $\phi$ as a configuration variable. 

Consider the infinite dimensional 
space of instantaneous
field configurations within the embedding $\iota$ to be $\mathcal{C}(\Sigma_\pm)\ni\phi_\mp$ depending on the 
three-surface they are defined on. We
 can therefore define a generalized $\mathcal{L}^2$-space for
the wave-functionals. The corresponding formal 
measure space is given by $\frak{M_\pm}=(\mathcal{C}(\Sigma_\pm),\mathcal{D}\phi^\mp)$ with 
infinite-dimensional uniform measure $\mathcal{D}\phi^\mp$. Let $\mathcal{L}^2(\frak{M}_\pm)$ denote 
the space of square-integrable, $\mathcal{D}\phi^\mp$-measurable 
wave-functionals $\Psi:\mathcal{C}(\Sigma_\pm)\to\mathbb{C}$ with semi-norm 
$\|\Psi\|_2=(\int_{\mathcal{C}(\Sigma_\pm)}\mathcal{D}\phi^\mp |\Psi|^2)^{1/2}<\infty$.
To define a proper norm, we need to divide out the wave-functionals yielding $\Psi[\phi]=0$
almost everywhere with respect to the functional measure \cite{hof19}. We note that
the functional measure is elected to be a uniform measure. 
As a basis for the wave-functional, we choose the eigenbasis of the field operator, such that all $\phi$
become multiplicative operators that yield the classical field $\varphi$ as an eigenvalue. 

To represent the momenta $\pi$ conjugate to $\phi$ 
in the configuration space, we must impose a consistent quantization prescription.
We will impose the quantization prescription in the Minkowski region where we also formulate our
initial conditions. 
For the dual-null formulation there exists a commonly used quantization that holds on every 
three surface $\Sigma_\pm$ as well as on $\mathcal{S}$ \cite{hei01}: 
\begin{eqnarray}
\left[\pi^\pm,\phi^\pm\right]_{\Sigma_\mp}&=&-i\delta^{(2)}(x,x')\delta(\xi_\pm,\xi'_\pm),\label{q1}\\
\left[\phi^+,\phi^-\right]_\mathcal{S}&=&-i\delta^{(2)}(x,x'),\label{q2}
\end{eqnarray}
where $\xi_+=v$ and $\xi_-=u$. Relations \eqref{q1} and \eqref{q2} suggest the representation
$\pi^\pm\to-i\tfrac{\delta}{\delta\phi^\pm}$. The resulting wave-functional for the free-field theory can 
be constructed by the following ansatz:  
\begin{equation}\label{psi}
\Psi[f](u,v)=N(u,v)\exp\left(-\frac{1}{2}[f]\mathcal{K}(u,v)[f]\right),
\end{equation}
with $\mathcal{C}(\Sigma_+)\times\mathcal{C}(\Sigma_-)\ni
f(x)=(\phi^+(x),\phi^-(x))^T$ as the field vector and the kernel matrix $\mathcal{K}_{AB}(u,v)$
where $A,B\in \{+,-\}$. The entries of $\mathcal{K}$ appearing in \eqref{psi} are bi-local functionals 
$\mathcal{K}:\mathcal{C}(\Sigma_\pm)\times\mathcal{C}(\Sigma_\pm)\to\mathbb{C}$, 
$(f_1,f_2)\mapsto[f_1]\mathcal{K}[f_2]$ of the 
form
\begin{equation}
[f]\mathcal{K}[f]=\!\!\!\iint_{\Sigma_{A,B'}}\!\!\!\!\!\!\!\!\!
\md \mbox{vol}_{A,B'}f^A(x)K_{AB}(x,x')f^B(x').
\end{equation}
It should be noted that the $f^A$'s are only defined on the corresponding $\Sigma_A$ while
the $x$-dependence in $K_{AB}$ has to be interpreted with respect to $\mathbb{W}^4$.
The primed index $B'$ signals that the three surface or volume integration respectively, is associated with the
primed coordinate. From here we see that $\mathcal{K}_{AB}$ develops a dependence on the $u$ and $v$ 
coordinates. The field independent part of \eqref{psi} also depends on $K_{AB}$ 
\begin{equation}\label{norm}
\frac{N(u,v)}{N(u_0,v_0)}=
\exp\!\left(\!-\frac{i}{2}\!\int^{u}_{u_0}\!\int^{v}_{v_0}\!
\int_\mathcal{S}\!\md^4x\sqrt{-g}\sum_{A,B}K_{AB}(x,x)\!\right)\!.
\end{equation}
A defining equation for the evolution kernel can be derived by plugging \eqref{psi} into the functional
Schr\"odinger equation and solving the resulting equation for the components $K_{AB}$. 

In full generality, the solution to these equations seems daunting, however, in a plane-wave background the 
high degree of symmetry simplifies them significantly. Extensive studies
\cite{gib75, garr91, neko17, sachs19} of these
space-times have shown that there occurs no mixing between outgoing and ingoing velocity fields. This 
leads to a decoupling of evolution directions and allows us to focus on the direction that crosses the plane wave, 
i.e. the ingoing fields $\phi^-$
traveling along the $u$-direction. Physically this makes sense, as we do not expect the $\phi^+$ fields 
evolving parallel to the wavefront to be affected by it. For the fields $\phi^-$,
we can formulate a well-defined initial value problem since we can use $\Sigma_-$ as an achronal set. 
Consequently, we perform a null-reduction, such that
we set the outgoing fields $\phi^+(v)\equiv0$ per default, then \eqref{fham} as well as \eqref{psi} will
depend on $u$ alone and the Schr\"odinger equation will reduce to $P=i\partial_u-2H(u)$ \cite{berg18}.
As we can see, \eqref{psi} simplifies such that we
can use a Gaussian ansatz in $\phi^-$ and the kernel matrix $\mathcal{K}_{AB}$
becomes a scalar bi-local function because $K_{- -}:=K$ will be the only non-vanishing contribution. 
In this specific situation, the resulting equation for $K$ is a Riccati equation that can be transformed to
the Klein-Gordon equation for the scalar modes $\varphi^-$ by inserting \cite{Jackiw,hof19}
\begin{equation}\label{ricctraf}
K_k(u)=\frac{-i}{\sqrt{-g}}\partial_u\ln\left(\frac{\varphi^-(u;k)}{\varphi^-(u_0;k)}\right).
\end{equation}
Fortunately, the exact solution to the mode equation on plane-wave space-times 
is known \cite{Friedlander1975}. 
For purely ingoing modes, the solutions can be constructed analytically using Huygens's
principle \cite{garr91,neko17}
to be of the form $\varphi_k^-(u,x)=\Omega(u)e^{i\phi_k}$ where $\Omega=|$det$(\gamma)|^{-1/4}$ and
\begin{equation}
\phi_k=\frac{k_0}{2}\Xi_{ab}x^ax^b+k_iE^i_ax^a+k_0v+\frac{F^{ij}k_ik_j}{2k_0}.\label{phase}
\end{equation}
Here, the shear-expansion tensor defined by $\Xi_{ab}={\dot{E}^i}_aE_{ib}$ describes the distortion 
along the $u$-direction after crossing the wave and the $k$'s come from a spatial Fourier
transformation. Employing the total differential d$\phi_k$ \cite{neko17},
we can deduce the $u$-derivative that 
appears in \eqref{ricctraf} to find 
\begin{equation}\label{ku}
K_{k}(u)=\frac{1}{\sqrt{-g}}
\left[\frac{k_0}{2}\dot{\Xi}_{ab}x^ax^b+k_i{\dot{E}^i}_ax^a+\frac{\gamma^{ij}k_ik_j}{2k_0}-
i\frac{\dot{\Omega}}{\Omega}\right].
\end{equation}
Note that $K$ develops a real and an imaginary part. With this in hand we are equipped to calculate the full 
wave-functional in order to study its behavior near the focusing-singularity.

\section{Shock wave} \label{IW}

Since we are interested in how the focusing singularity affects ingoing modes 
after passing through the wave, we consider the example of a shock wave
$H_{ab}(u)=\lambda^{(a)}\delta(u)\delta_{ab}$ at $u=0$, described by a Dirac delta distribution.
Here, $\lambda^{(a)}$ is the eigenvalue that denotes the physical amplitude 
of the wave in the $x^a$-direction. 
\subsection{Brinkmann coordinates}\label{BC}
For this
analysis we will start with Brinkmann coordinates because of the wave's explicit appearance in the metric. 
 The focusing singularity can be shown to occur 
at a distance $u_{\rm f}$ from the wave-front that is inversely proportional to $
\lambda^{(a)}$. To see this, we first calculate the 
explicit form of the vierbeins by solving $\ddot{E}_{ia}=H_{ab}E^b_i=\lambda^{(a)}\delta(u)E_{ia}$ 
with boundary conditions at $\mathscr{I}^\pm$: $\lim_{u\to\pm\infty}{E^\pm}^a_{i}={\delta^a}_i$. We
find for the shock wave with the corresponding boundary condition
\begin{equation}
E^\pm_{ia}=\delta_{ia}\left(1\mp\lambda^{(a)}u \Theta(\mp u)\right),\label{4bein}
\end{equation}
$\Theta(u)$ being the Heaviside theta-distribution. Insertion of \eqref{4bein}
into \eqref{rosen} shows that the metric 
becomes degenerate at the value $u^\pm_{\rm f}=\mp1/\lambda^{(a)}$ as stated before. The pair 
of focusing points is due to the past or future evolution. In the following, we will restrict ourselves to 
only the future-development of the ingoing fields and suppress the superscript of $E^-_{ia}$
(we choose to start at $\mathscr{I}^-$ which fixes the boundary condition there).
Because we have chosen a symmetric setup by setting
the shock wave to $u=0$, we could easily carry out the calculation for the past development by choosing 
the opposite sign. In order to calculate the explicit form of $K(x,x')$, we need 
the metric $\gamma_{ij}=\delta_{ij}(1+\lambda^{(i)}u \Theta(u))(1+\lambda^{(j)}u \Theta(u))$ and its 
determinant
$\gamma=(1+\lambda^{(1)}u \Theta(u))^2(1+\lambda^{(2)}u \Theta(u))^2$. The last relevant 
contribution in \eqref{phase} is the shear-expansion tensor 
\begin{equation}\label{xiab}
\Xi_{ab}=\delta_{ab}
\frac{-\lambda^{(a)}\Theta(u)(1+\lambda^{(b)}u\Theta(u))}{(1+\lambda^{(a)}u \Theta(u))^2}.
\end{equation}
Without loss of generality, we omit the second term appearing from the differentiation 
which is proportional to $u\delta(u)$ and hence vanishes identically for every value of $u$.
The tensor in \eqref{xiab} describes distortion of geodesics and consequently
the focusing at $u_{\rm f}$ for $\lambda^{(a)}<0$. Such an eigenvalue will definitely
exist for gravitational waves since $H(u,x)$ is traceless. The shear-expansion tensor manifests a memory
effect, that is, it is zero unless the field has encountered the plane wave. In other words, by looking at this
tensor we can determine immediately whether or not a scalar field has met a plane wave or not. 
As we can see, $\gamma^{ij}$ diverges for
$u\to u_{\rm f}$ as well as $\Xi_{ab}$, $\Omega$, and $E^i_a$ while \eqref{4bein} and $\gamma_{ij}$ 
approach zero. The $u$-derivative of $\Xi_{ab}$ is then given by
\begin{equation}\label{xidot}
\dot{\Xi}_{ab}=\delta_{ab}\left[
\frac{(\lambda^{(a)}(\lambda^{(b)}u+2)-\lambda^{(b)})\lambda^{(a)}\Theta(u)}
{(1+\lambda^{(a)}u \Theta(u))^3}+
h(u)\right].
\end{equation}
Besides the singular contribution at the focusing point, \eqref{xidot} will also develop a divergence at
$u=0$ because the second term $h(u)$ is just proportional to the shock wave
 $H_{ab}(u)\propto\delta(u)$ for all $u\in \mathbb{R}$.
This divergence is not physical however, as it simply stems from the choice of $H(u,x)$ as delta distribution. 
As we are not interested in the wave itself, we will set this term to zero due to the vanishing support at 
$u\neq0$. For more realistic smooth profile functions it remains finite in any case.

The functional Schr\"odinger representation is particularly useful to studying the influence of space-time
singularities on quantum field theory. Formerly, the concept of quantum completeness 
\cite{hof15,hof17,egl17,hof19} has been used to investigate the Schwarzschild and Kasner singularity.
Those space-times admit a globally hyperbolic slicing in which the singularity is not a part of the 
manifold. In fact, we could see it as a geodesic border towards which quantum fields could leak into the 
classically singular configuration.
Structurally this type of singularity is different from the null-singularity in plane-wave
space-times. The former space-times admit a spacelike singularity where geodesics (and so space-time itself) 
end abruptly. In our case, the space-time itself is perfectly regular at the focusing point if one considers
the fully extended space-time in Brinkmann coordinates. However, $(\mathbb{W}^4,g)$ is null-geodesically
incomplete because of the focusing on $\Sigma_{{\rm f}-}$. Due to the intersection of geodesics we lose 
predictability but we could still cross $\Sigma_F$ in Brinkmann coordinates. In Einstein-Rosen
coordinates, we would instead find that at the focusing point the Einstein-Rosen patches 
$(\mathbb{W}^\pm,g^\pm)$ end abruptly. While there is no analytic 
extension in Schwarzschild and Kasner space-times, we must adapt the scientific objective to the situation
where space-time continues beyond the singular hypersurface. Hence, for the presented setup the research
question should be whether or not quantum field theory develops pathologies while crossing the focusing
hyperplane. In other words we investigate whether the wave-functionals remain normalizable throughout
the whole evolution such that $\|\Psi\|_2^2<\infty$, $\forall u$ on $(\mathbb{W}^4,g)$. 

Taking \eqref{4bein}, we can derive all relevant contributions that arise in \eqref{ku} where we can
identify the problematic contribution to be the first term in \eqref{xidot}. We choose a linearly
polarized plane-wave \cite{sachs19} with $\lambda^{(1)}=1$ and $\lambda^{(2)}=-1$, or equivalently, 
$H_{ab}(u)=\sigma^z_{ab}\delta(u)$, where $\sigma^z_{ab}=\text{diag}(1,-1)$. In this system,
the focusing occurs at $u_{\rm f}=1$; the problematic value is the case of the negative eigenvalue. 
This choice greatly simplifies \eqref{xidot} since we are only interested in the region close to the singularity, i.e.
$u\approx1$ where $\Theta(u_{\rm f}\pm\Delta u)=1$ for $0<\Delta u<1$, such that the second term in
\eqref{xidot} stays zero. These assumptions simplify \eqref{xidot} further
such that the leading singularity is given by
$\dot{\Xi}_{22}\to 1/(1-u)^2$ which is clearly divergent at the focusing point. In fact, all other
contributions in the real part of \eqref{ku} scale similarly. We use our knowledge to determine the behavior 
of $\Psi[\phi^-](u)$, first for the field independent part. It is worth pointing out that the field independent part of 
the wave functional $N(u)$ is usually referred to as the normalization due to its resemblance to a 
normalization constant in quantum mechanics. Due to its time-dependence, however, it can not be chosen to 
compensate the norm of the field-dependent part at all times so as to ``normalize'' the functional.
The resulting state would clearly fail to solve the Schr\"odinger equation for 
later times. We are interested in the norm of $\eqref{psi}$ in close proximity to $u_{\rm f}$
and therefore only consider the most divergent terms. The normalization $|N|^2(u)$ is determined by the
imaginary part of \eqref{ku}
\begin{equation}
\frac{\dot{\Omega}}{\Omega}=
\frac{-(\lambda^{(1)}+\lambda^{(2)}+2\lambda^{(1)}\lambda^{(2)}u)\Theta(u)}
{2((1+\lambda^{(1)}u\Theta(u))(1+\lambda^{(2)}u\Theta(u)))},
\end{equation}
where we use  $\Omega=|(1+\lambda^{(1)}u\Theta(u))(1+\lambda^{(2)}u\Theta(u))|^{-1/2}$. 
The expression can be simplified by assuming $\lambda^{(1)}=1$, $\lambda^{(2)}=-1$,
and $u>0$ to be $\dot{\Omega}/\Omega=u/(1-u^2)$.
As we see, this term is divergent in the limit $u\to 1$, even after performing the 
$u$-integration in \eqref{norm} which yields $\ln(1-u^2)$ in the exponent.
We then find for $0<u_0<1$
\begin{equation}
\frac{|N(u)|^2}{|N(u_0)|^2}=|1-u^2|^{{\rm vol}(\Sigma_-)}\label{norm}
\end{equation}
where the volume factor vol$(\Sigma_-)$ comes from the integration over the 
hypersurface $\Sigma_-$ in \eqref{norm} and can be interpreted as an infrared regulator. Due
to the Heaviside distribution, $N(u)$ is trivially equal to one for all $u<0$, as in this case space is just a 
Minkowski patch.

Similarly, we can address the field-dependent part. First, we note that only the real part of 
\eqref{ku} contributes. The determinant $g$ in the Brinkmann chart is just given by $-1$ and is hence not 
$u$-dependent. Altogether, we find that
\begin{equation}\label{exp}
\left\|\exp\left(-[\phi^-]\mathcal{K}[\phi^-]\right)\right\|^2_2=
\sqrt{\left|\frac{C}{{\rm Det}(-g{\rm Re}(K))}\right|}
\propto|1-u|^{\tfrac{\Lambda}{2}}
\end{equation}
where the last expression is the leading behavior near the focusing singularity $u_{\rm f}=1$. We have 
absorbed all
constants into $C$ and introduced 
the ultraviolet cut-off $\Lambda$ that regulates the infinite dimensionality
of the field space. We will comment on this in more detail later.
It should be noted that because the
space-time itself does not terminate at the focusing singularity, we need to comment on the evolution 
beyond
$u_{\rm f}$. As it can be checked by the explicit form of \eqref{ku}, the kernel function resembles a 
Minkowski limit for $u\to-\infty$ such that $\Psi[\phi^-](u)\to \Psi_{\mathbb{M}^4}[\phi^-]$ with 
$\Psi_{\mathbb{M}^4}[\phi^-]=N_0\exp(-[\phi^-]\mathcal{K}_{\mathbb{M}^4}[\phi^-])$ being the Minkowski
wave-functional and $\mathcal{K}_{\mathbb{M}^4}$ the square root of the 
inverse Minkowski propagator in the dual-null foliation. In the limit $u\to\infty$, we see that the terms
dominating at $u_{\rm f}$ are subdominant and \eqref{norm} as well as \eqref{exp} approach constant 
values.
Although the terms tend asymptotically to a Minkowski solution, the contributions from $\Xi_{ab}$ will not vanish which
accounts for the gravitational memory of those solutions that have crossed the gravitational wave.
\begin{figure}
\includegraphics[width=8.6cm]{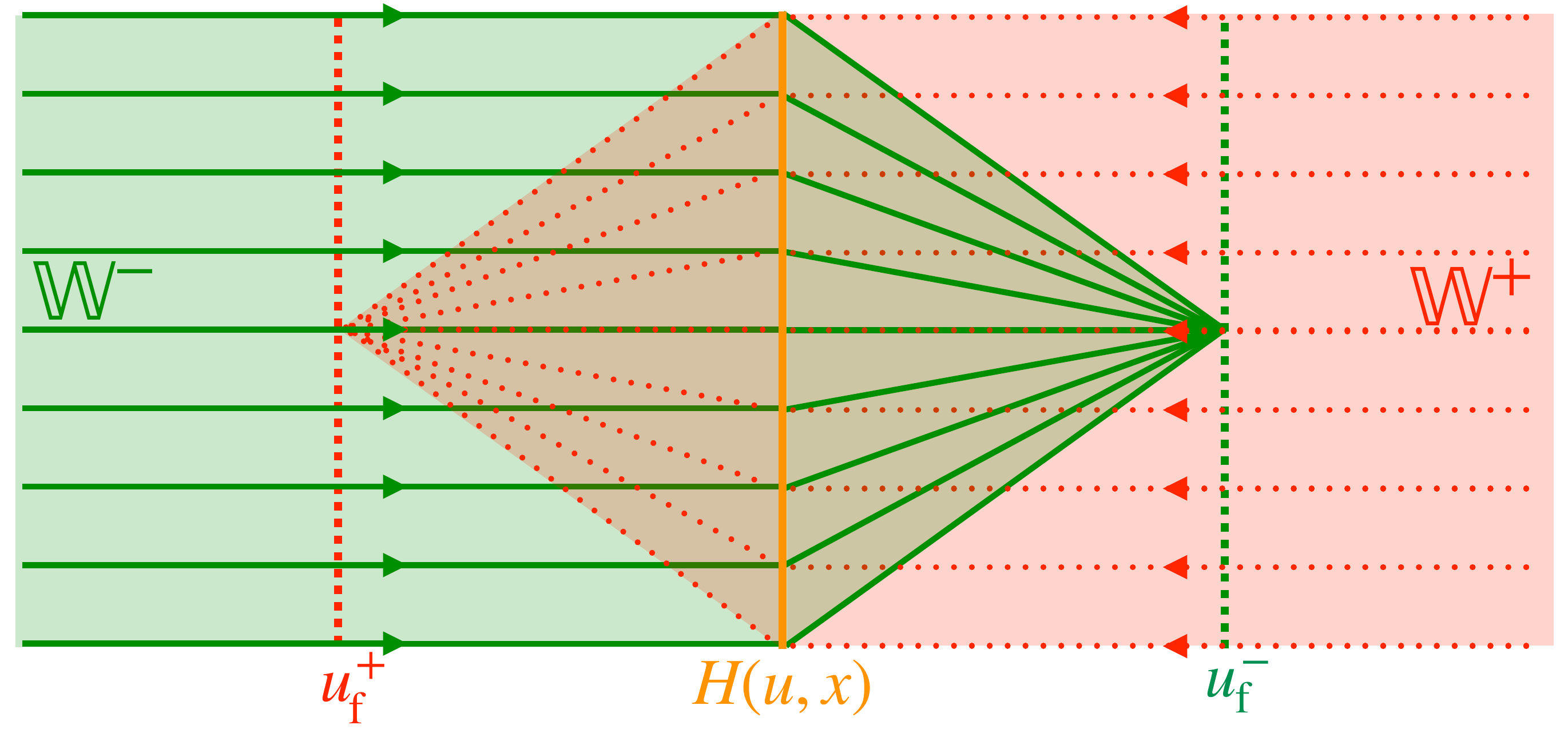}
\caption{
\label{fig:plot1} 
Focusing that occurs in the different Einstein-Rosen patches $\mathbb{W}^-$ and $\mathbb{W}^+$ is
described by the green and red lines. Einstein-Rosen coordinates do not extend beyond the null singularities
such that $\mathbb{W}^-$ is given by the green shaded and $\mathbb{W}^+$ by the red shaded area.
The diamond-shaped overlap belongs to both patches. Brinkmann patches $\mathbb{W}^4$
have non-degenerate focal planes and the vol$(\Sigma_{{\rm f}^-})$ is non-zero in contrast to
vol$(\Sigma_{{\rm f}^-})$ in Einstein-Rosen coordinates at the null-singularities.}
\end{figure}
\subsection{Einstein-Rosen coordinates}\label{ERC}
From the perspective of Einstein-Rosen
coordinates, the setup is similar to the Kasner or Schwarzschild case where the singularity marks the 
geodesic border. However, as we see in FIG. \ref{fig:plot1}, the degenerate hypersurface 
$\Sigma_{u_{\rm}^-}$ with the focusing singularity on the patch $(\mathbb{W}^-,g^-)$ is just a 
Minkowski hypersurface for $(\mathbb{W}^+,g^+)$, the other Einstein-Rosen patch. Performing the 
diffeomorphism $\{u,v,x^a\}\to\{U,V,y^i\}$, we can deduce the kernel in $(\mathbb{W}^-,g^-)$ to be
\begin{equation}\label{kuer}
K(U)=\frac{1}{\sqrt{-\gamma}}
\left[\frac{\gamma^{ij}k_ik_j}{2k_0}-i\frac{\dot{\Omega}}{\Omega}\right].
\end{equation}
In the limit of large $U$ 
values it reduces precisely to the kernel for a Minkowski functional \cite{hat18}, and while approaching the focal 
point it generates an imaginary part, as well as a non-trivial $U$-dependence. Calculating $\|\Psi\|_2^2(U)$ in 
this coordinate neighborhood, yields the same result as \eqref{norm}
for the normalization. However, the volume of the singular hypersurface vol$(\Sigma_-)$
is in principle infinite in Brinkmann coordinates. Since none of the $x^a$-directions degenerate, the 
volume of a Brinkmann sub-manifold is equal to a sub-manifold of a dual-null foliated
Minkowski space-time. However, for the Einstein-Rosen metric, the volume vol$(\Sigma_-)$ tends to $0$ because
the sub-manifold degenerates in, at least, one of the $y^i$-coordinates; this can be seen from the 
form of the vierbeins at the focal point. Hence, in Einstein-Rosen coordinates, $|N(U)|^2\to1$.

The field-dependent part again requires the regularization of the functional determinant similar to \eqref{exp} For Det$(-\gamma$Re$(K))$ and \eqref{kuer} 
we find by explicitly writing out the eigenvalues that
\begin{equation}\label{det}
\mbox{Det}(-\gamma\mbox{Re}(K))=\prod_{k_1,k_2}
\left(\frac{1+U}{1-U}k_1^2+\frac{1-U}{1+U}k_2^2\right).
\end{equation}
Again this bears great resemblance to the Minkowski result. We can now employ a zeta function regularization 
in order to obtain a finite result and extract from it the relevant $U$-behavior. A very similar calculation was 
already carried out in the context of the generating functional of Thirring models \cite{SachsWipf} and may be 
used with slight alterations. To do so we rewrite the infinite product over the 
eigenvalues $\lambda_k$ in the following form
\begin{equation}\label{prod}
	\prod_{k}\lambda_{k}=\!\! \prod_{n\in \mathbb{Z}^2}\!\!\!\left(\frac{2\pi}{L}\right)^2 \!\!\!\rho^{\mu\nu}\!\left(\tfrac{1}{2}\mathds{1}_\mu\!+c_\mu\!+n_\mu\right)\!\!\left(\tfrac{1}{2}\mathds{1}_\nu\!+c_\nu\!+n_\nu\right),
\end{equation}
by choosing toroidal compactification of length $L$ to regulate the infinite volume for now. 
Here, $n_\mu$'s are reference coordinates on the torus, $\rho$ is the reference metric,
$c_\mu$ is a constant shift, and 
$\mathds{1}_\mu=(1,1)^T$ is the one-vector. Equation \eqref{prod}
can be re-expressed in terms of a generalized $\zeta$-function
\begin{equation}
	\zeta(s):=\sum_{k}(\lambda_{k})^{-s}.
\end{equation} The desired determinant will then be given by the value of the derivative of the zeta-function 
at $s=0$. Writing the zeta function in terms of a Mellin transform and using the Poisson re-summation 
formula, we arrive at
\begin{align}
	\zeta(s)=\frac{\Gamma(s-1)}{\Gamma(s)}\pi^{2s-1}&\sqrt{\rho}\sum_n(\rho_{\mu\nu}n^\mu n^\nu )^{\frac{s-1}{2}} \nonumber\\&\times\exp\left[-2\pi n^\mu\left(c_\mu+\tfrac 1 2\mathds{1}_\mu\right)\right].
\end{align} Fortunately due to the form of \eqref{det} we read off $c_\mu=-\tfrac12\mathds{1}_\mu$
such that the phase factor is equal to unity; $\rho=
\mbox{diag}(\tfrac{1+U}{1-U},\tfrac{1-U}{1+U})$ and the derivative at zero is given by 
\begin{equation}
\zeta'(0)=\pi^{-1}\sqrt{\rho}\sum_n(\rho_{\mu\nu}n^\mu n^\nu )^{-\frac12}.
\end{equation} The sum is well-defined in terms of the Epstein zeta-function and gives a result independent of 
$U$. The technical reason for this is that the determinant of the metric in \eqref{det} is constant. In 
\cite{hawk77}, Hawking gives a slightly different argument, noting that the integrals depending on the
infra-red regulator must vanish, confirming that the determinant will become constant.
In the Appendix, we show that the zeta-regularized determinant in
\eqref{det} will just become a constant that can be set to one. 
In Einstein-Rosen coordinates the volume element $\text{vol}(\Sigma_-)$ of the transversal 
directions appearing in the expression for the normalization factor \eqref{norm} degenerates with respect
to one spatial direction as seen from \eqref{4bein} explicitly. While for Brinkmann the transversal sub-manifold 
$\Sigma_-=\mathscr{V}\times\mathcal{S}$ is a flat three-manifold like in a dual-null foliated Minkowski 
space-time, the spatial part $\mathcal{S}$ degenerates in Einstein-Rosen coordinates. Suppose we start
with a spatially limited bundle of rays with a quadratic cross-section at $\mathscr{I}^-$. After traversing
the wave, the cross-section will elongate in the $x_1$- and contract in the $x_2$-direction. Albeit the elongation
is finite (here it is doubled), the contraction will be total such that the cross-section degenerates.
Hence, the volume vol$(\Sigma_-)$ essentially captures a subtle dependence on $U$ such that one
$y$-coordinate degenerates and the resulting volume shrinks to zero when $\Sigma_-$ becomes a 
null two-manifold:
\begin{equation}
\lim_{U\to 1}|1-U^2|^{{\rm vol}(\Sigma_-)}=1\quad\mbox{with}\quad\lim_{U\to1}\mbox{vol}(\Sigma_-)=0.
\end{equation} 
Therefore, the contribution to the overall probability stemming from the field-independent part of the wave-
functional will simply equal one at the focal point. As we have seen prior, the functional determinant will 
approach a constant too, that may be set to one by an appropriate choice of the constant $C$. Thus, we see 
that the wave-functional remains normalizable throughout the entire evolution and its norm 
$\| \Psi\|^2_2(U)\to1$ at the focal point.

The results in \ref{BC} and \ref{ERC} seem to be in tension as the probabilities do not agree. This effect
is due to the gravitational memory. 
Unlike the examples studied in \cite{hof15,hof19}, the geometry does not trivialize the theory 
and the singular configuration is populated, as the wave-functional enjoys probabilistic support. 
Moreover, it will be reached inevitably by the essentially Minkowski-like evolution,
resulting in a probability of exactly one at the geodesic border.
Due to the fact that Brinkmann coordinates are a non-trivial extension of the Einstein-Rosen metric,
a direct comparison is intricate. This is due to the difference in the functional spaces over which the functional 
integration is understood in \eqref{exp}. The solution space to the Brinkmann d'Alembert operator $P_B$ is a 
unification of the solution spaces associated with both Einstein-Rosen patches.
 In other words, as the Brinkmann patch is an extension of 
the Einstein-Rosen patch, we are also integrating over modes that lack an interaction with the plane
wave in the past as illustrated by FIG. \ref{fig:plot1}. As can be checked explicitly, the solution
space to $P_B\phi=0$ consists of solutions that show a memory effect since they have passed the wave
and solutions that are plane waves. Considering the explicit form of the mode solutions for $u>0$,
\eqref{phase} describes two different kinds of modes in Brinkmann space-time: one kind corresponds to the
(green) $\mathbb{W}^-$ patch and show a gravitational memory because of the non-trivial vierbeins
$E_{ia}^-=\delta_{ia}(1+\lambda^{(a)}U)$,
 while the other modes belong to the (red) $\mathbb{W}^+$ patch and have 
$E_{ia}^+\equiv\delta_{ia}$, that is to say, they have an empty gravitational memory in the future development.

The inclusion or omission of these modes will naturally affect the induced probability measure. For a direct 
comparison consider a finite region in the patch where both coordinates overlap, i. e. the right
part of the diamond. The functional measure in Brinkmann space-time
will not distinguish between the $\mathbb{W}^-$
modes and the $\mathbb{W}^+$ modes and an integration on a hypersurface will 
therefore incorporate both modes. To make it comparable to an Einstein-Rosen development, we need to 
project out modes with empty gravitational measurement in order to match both results. The remaining 
modes with a non-empty gravitational memory will be confined within a region (right side of the diamond)
that, to a Brinkmann observer, will continuously shrink until it ends at a caustic, thereby rendering the evolution the same as for the local Einstein-Rosen observer. 

\section{Back-reaction}

Ending inevitably in the coincidence limit it is clear that the evolution requires some form of completion to remain physically viable. Classically, the energy momentum tensor will diverge at the focal point and it is therefore to be expected 
that back-reactions may regularize the singular behavior, at least to some degree. While exact statements are difficult to come by, owing to the technical difficulty of the resulting expressions, some general observations can be made. It should be mentioned that we assume we will stay in the class of plane-wave space-times that is, all perturbations of the wave-front itself are ignored
throughout this analysis. To get an estimate for the leading contribution 
we focus on the strongest diverging component of the energy momentum tensor \cite{sachs19}
\begin{equation}
	T_{uu}=(\partial_u\varphi^-)^2\approx\left(\partial_u\sqrt{|E|}^{-1}\right)^2\sim
	\frac{\dot{\mathcal{E}}^2}{\mathcal{E}^3},
\end{equation}
where we defined $\mathcal{E}:=|\mbox{det}(E)|$. 
Since our aim is to calculate a back-reacted metric from the 
Einstein equation, we use our knowledge that the only non-vanishing contribution to the Ricci-tensor will come 
from the $uu$-component and that in our case the Ricci scalar is zero, to find the Einstein tensor's 
$uu$-component
\begin{equation}
	G_{uu}=R_{uu}=\frac{\ddot{\mathcal{E}}}{\mathcal{E}}.
\end{equation}
Note that our ansatz for the
back-reacted metric is of Einstein-Rosen form such that our objective is to construct
an improved $\gamma_{ij}$. We may attempt to find a 
self-consistent solution to $G_{uu}=T_{uu}$ in terms of the tetrad $E$. 
In this case the solution for the square root of the determinant is given by
\begin{equation}\label{correctedtetrad}
	\mathcal{E}=\frac{1}{\ln(1-u)},
\end{equation}
to leading order as $u$ approaches one. We can construct a new
$\gamma$ as a diagonal matrix by treating \eqref{correctedtetrad} 
as the tetrad that corresponds to the degenerating entry while we set the remaining entry to be
one for simplicity, as it would tend toward a constant anyway. This yields the back-reacted metric 
\begin{equation}
\gamma=
\left( {\begin{array}{cc}
		1 & 0 \\
		0 & \frac{1}{\ln^2(1-u)} \\
\end{array} } \right).
\end{equation}
In this approximate case, the expression for the shear tensor becomes very straightforward, in fact it 
reduces to just one component as there is only one non-vanishing component of $\dot{E}$,
\begin{equation}
	\dot{\Xi}_{22}=\frac{1-\ln(1-u)}{(1-u)^2\ln^2(1-u)}.
\end{equation}
This is still divergent as $u$ approaches 1, but weakened through the logarithmic factor in the denominator.
Approximating $K(u)\sim\dot{\Xi}_{22}$ shows that Det$(-g$Re$(K))^{-1/2}$ decreases
slower than \eqref{exp} but eventually approaches zero as well.
The same holds for the imaginary part of the Kernel
\begin{equation}
	\frac{\dot\Omega}{\Omega}=-\frac{1}{2(1-u)\ln(1-u)}.
\end{equation} 
The qualitative behavior is thus unchanged, however the strength of the contraction near the focal point is 
attenuated. This feature seems to be systemic unless a disturbance in $v$-direction of the wave-front 
is considered. In this case, we leave the class of plane-wave space-times giving $H(u,v,x)$ an explicit
$v$-dependence that induces a mode mixing between $\phi^+$ and $\phi^-$ which intuitively leads to a 
de-focusing. However, a singularity avoidance might only result from a much less idealized system since
a na\"ive addition of a $v$-dependence might result in a Khan-Penrose space-time \cite{khan}
where the singularity becomes spacelike.

\section{Conclusion} 
This article investigates the evolution and consistency of massless quantum fields in a plane-wave space-time. 
We constructed the functional Schr\"odinger formalism in the dual-null embedding such that
we can formulate a well-defined initial condition and boundary value problem along the Hamiltonian flow.
On its own, this formalism has various applications, e.g. it can be used to study systems where evolution on 
null paths is favored, such as gravitational tunneling across horizons \cite{gia20}. One objective of this
article was to test the hypothesis of incompleteness by using quantum fields as probing devices. We found indeed
that the null-singularity will be populated by quanta that have passed the wave. This is not surprising since 
the focusing of the plane-wave is not fronted by any effect that can prevent the caustic since, here,
the space-time itself is Minkowski and the field theory is free. In other words, we saw that unitarity of the 
dual-null evolution forces quantum fields into the caustic while keeping normalizability intact. 

An interesting feature of these results comes from the comparison between the two different patches: while in
Einstein-Rosen coordinates, we recovered, that within one patch, there exists a unitary evolution for all initial
conditions and the probability is conserved throughout the evolution, Brinkmann coordinates show a 
steady decrease of the probability amplitude. Brinkmann patches are
non-trivial extensions of Einstein-Rosen patches with respect to field configurations. Geometrically, the Brinkmann patch
consists of Minkowski hypersurfaces that cover both Einstein-Rosen patches. This in turn implies for the field configurations that the 
Brinkmann configuration space $\mathcal{C}^{\mathbb{W}^4}\corresponds\mathcal{C}(\Sigma_u)$
is the combination of the configuration spaces of both Einstein-Rosen patches 
$\mathcal{C}^{\mathbb{W}^\pm}\corresponds\mathcal{C}(\Sigma_U^\pm)$ 
with hypersurfaces $\Sigma_u$ and
$\Sigma_U^\pm$ in Brinkmann geometry, and past/future geodesically incomplete Einstein-Rosen geometries
respectively. 
Hence, the functional measure space for Brinkmann 
$\frak{M}^{\mathbb{W}^4}$ equals $(\mathcal{C}(\Sigma_u),\mathcal{D}\phi_{\mathbb{W}^4})$ which in turn equals $
\frak{M}^{\mathbb{W}^+}\cup\frak{M}^{\mathbb{W}^-}$, where $\phi_{\mathbb{W}^4}$'s are the
instantaneous field configurations on a Brinkmann null hypersurface. At the null-singularity $\frak{M}^{\mathbb{W}^-}$ degenerates to a null set with respect to 
$\frak{M}^{\mathbb{W}^4}$ which inevitably entails the vanishing of the probability amplitude for these
configurations as they have measure zero according to $\mathcal{D}\phi_{\mathbb{W}^4}$. In other 
words, the Brinkmann observer would describe a focusing of the fields into a caustic for gravitational waves.
Albeit the results for the norm of the wave-functionals
seem to develop a tension at first glance, they are in agreement when properly
considering the subtleties of the configuration spaces. As we saw, various subtleties arise from treating such 
infinite dimensional measures that need due care. Situations like a quantum probing involve free fields described by Gaussian ground states
which are tractable and lead to a well-defined measure. However, non-Gaussian deformations can be treated through a (time-dependent)
Rayleigh-Schr\"odinger perturbation theory \cite{hof17}.
Since the functional Schr\"odinger equation describes the evolution of
the field configurations itself, there is limited information we can extract concerning the individual degrees of freedom.

It is always difficult to reflect on the physical nature of singular solutions in general relativity. One of the more general statements known from globally hyperbolic space-times, such as Kasner universes, is the Belinskii-Khalatnikov-Lifshitz (BKL) conjecture, claiming that in the approach to the singular point, velocity terms will dominate 
over the potential terms $\dot{\varphi}\ll\varphi'$ \cite{BKL1,berg93}. Although the original formulation is tied to spacelike singularities, a similar BKL behavior has been observed for the null-singularity at inner horizons in Kerr black holes \cite{ori99}. In the present example, we consider the derivative 
of the scalar field $\md\varphi=(\md\Omega+\Omega \md\phi_k)e^{i\phi_k}$,
where the $\md\Omega$-term
only arises for the $u$-derivative. To estimate the behavior close to the focal point, we use 
\eqref{phase}, the explicit form of $\dot{\Omega}$, and $\partial_a\phi_k=k_iE^i_a+k_0\Xi_{ab}x^b$,
the spatial derivative of the phase. The scaling shows that the kinetic term will go as $\dot{\varphi}\sim 
\dot{\Omega}+\Omega\dot{\Xi}\sim1/(1-u)^{5/2}$ while the spatial derivative will be
$\varphi'\sim\Omega\Xi\sim1/(1-u)^{3/2}$
in leading order.

Our analysis shows explicitly that, although the equations of motions might suggest a BKL-type behavior, the
wave-functional in plane-wave space-times
shows major differences to the wave-functional in a Kasner universe
\cite{hof19} and that of a Schwarzschild space-time \cite{hof15}. Most notably, the probability density in these previously examined cases decreases to zero towards singularity. 
The explanation of this difference in qualitative behavior lies in the physics of the underlying space-time:
plane-wave space-times provide no curvature that can act as a source for dissipation such as in Kasner or Schwarzschild cases, 
i.e. they provide a unitary evolution in the Schr\"odinger picture while this cannot hold on a time-dependent,
curved manifold like Kasner or Schwarzschild \cite{tor99,ag15}. In this case, information from the quantum sector gets transferred to the classical background \cite{egl17} leading to the question how one can read all pieces of information in this representation.

 Another distinctive feature of both of these examples, is the pronounced anisotropy developed by the singular hypersurface. In both cases at least one of the spatial directions grows without bound as the others contract and thus complete coincidence is avoided. In the present example this is not the case, as all spatial directions are either contracting or bounded and therefore nothing prevents a coincidence at the caustic.

In conclusion, we see that the caustic singularity may only be avoided when the strict symmetry requirements on the space-time are lifted. This may be achieved by the possibility of mixing the $u$ and $v$-dependency of the modes, either by allowing for the field to have self-interactions or alternatively, by considering back-reactions on the space-time, leading to less restrictive geometries.

\begin{acknowledgments}

We thank Maximilian K\"ogler for his comments on the manuscript and
Aditya Singh Mehra for helpful suggestions concerning the null-reduction of the 
Schr\"odinger equation. We also thank Stefan Hofmann for insightful discussions on topic of the BKL conjecture. Our special thanks go to Sergio Zerbini for the fruitful discussion on the
zeta-function regularization and related subtleties caused by massless fields. This work was partly funded by the Excellence Cluster Origins of the DFG under Germany’s Excellence Strategy EXC-2094 390783311. We appreciate financial support from the Alexander-von-Humboldt foundation. 

\end{acknowledgments}
 
\appendix 
\section{Zeta-function regularization}\label{app:zeta}

Integration over the space of field configurations as performed in \eqref{exp} leads to functional determinants
that have to be regularized due to the infinite dimensionality of the configuration space 
$\mathcal{C}(\Sigma_-)$.  A standard technique (cf. \cite{hawk77,eli98,zerb02,zerb04,mor11})
 involves a generalized local zeta-function of
some operator $\mathcal{O}$ \cite{mor11}
\begin{equation}
\begin{aligned}
\zeta\left(s;x,y|\tfrac{\mathcal{O}}{\nu^2}\right)=&\int_0^\infty\!\!\!\!\md \tau
\frac{\nu^{2s}\tau^{s-2}}{\Gamma(s)}\left[L_\tau(x,y|\mathcal{O})-P_0(x,y|\mathcal{O})\right]\\
=& \sum_j\left(\frac{\lambda_j}{\nu^2}\right)^{-s}\!\!\!\!\!\phi_j(x)\phi^*_j(x),\label{zeta}
\end{aligned}
\end{equation}
where $L_\tau(x,x)$ represents the kernel of the operator, $P_0(x,x)$ are the zero-modes, and
$\nu$ is an arbitrary scale which we set to one.
The second equation clarifies the relation to functional determinants by introducing
the eigenvalues $\lambda_j$ corresponding to the eigenvectors $\phi_j(x)$ that are normalized as 
$\textstyle{\int}$d$^4x\sqrt{-g}\phi_a(x)\phi_b(x)=\delta_{ab}$ with respect
to the background \cite{hawk77}.
The convergence of \eqref{zeta} in two dimensions is ensured whenever Re$(s)>1$.
The functional determinant is then equivalent to
\begin{equation}\label{func}
\mbox{Det}(\mathcal{O})=\prod_j^N \lambda_j=
\left.e^{-\tfrac{{\rm d}}{{\rm d} s}\zeta(s;x,y|\mathcal{O})}\right|_{s=0}.
\end{equation}
Following \cite{hawk77,eli98}, we use the explicit form of \eqref{ku}
in Einstein-Rosen coordinates and reformulate the determinant using \eqref{func} to define the 
zeta function
\begin{equation}
\zeta(s;U|K)=\frac{\rm Vol}{4\pi^2}\int\frac{\md^2k}{(2\pi)^2}
\left(\frac{1+U}{1-U}k_1^2+\frac{1-U}{1+U}k_2^2\right)^{-s},
\end{equation}
where $k_i$ are the momenta in the directions $y^i$ and Vol is a volume regulator. The above integral
can be simplified by a transformation of the measures $\md k_1\to f(U)\md \ell_1$ and 
$\md k_2\to f^{-1}(U)\md \ell_2$ where we introduced $f(U)=\sqrt{(1+U)/(1-U)}$. It can be seen
that the prefactors from the Jacobi determinants cancel each other and 
the $U$-dependence vanishes. After
shifting to polar coordinates $\{\ell,\alpha\}$, the angular $\alpha$-integration just yields a constant and
we are left with an integral that can
be analytically continued such that \cite{gelfand,zerb02}
\begin{equation}
\int_0^\infty\md\ell \ell^z=0,
\end{equation}
from which we get that $\zeta(s;U|K)|_{s=0}$ is identically zero and so is the derivative.
For the functional determinant in Einstein-Rosen coordinates it follows immediately that
Det$(-\gamma$Re$(K))=\exp(-\tfrac{\rm d}{{\rm d}s}\zeta(s;U|K))|_{s=0}=1$
for all $U\in(0,1]$.
In \cite{hawk77}, Hawking gave a heuristic argument by introducing an infrared regulator $\varepsilon$ and
found that replacing the lower integration bound by $\varepsilon$, the result yields $\varepsilon^{2-2s}$
(for our dimensions) which goes to zero in the infrared limit.

\bibliography{nullliteratur}
\end{document}